# Sustained Coherence Characteristics of Failure Process of Shallow Buried Tunnel under Multiple Explosive Loads based on Persistent Homology

Shengdong ZHANG[1]   Shihui YOU*[1]   Longfei CHEN[2]   Xiaofei LIU[2]

1. College of Mechanical and Electrical Engineering, Zaozhuang University 277160;
2. College of Civil Engineering and Mechanics, Xiangtan University 411105

**Abstract:** The damage characteristics of a shallow buried tunnel under multiple explosive loads is an important research issue in the design and evaluation of protective engineering. It is of great significance to develop a method for early warning of the safety of the shallow buried features. The discrete element method is used to establish a mechanical model of the shallow buried tunnel. The South Load Equivalent Principle treats blast loads as a series of dynamic forces acting uniformly on the surface. Based on the discrete element method, the dynamic response after each blast load and the damage evolution process of the surrounding rock of the tunnel are obtained. The strength reduction method is used to obtain the surrounding rock of the tunnel. Introduce the theory of continuous homology, and use the mathematical method of continuous homology to quantitatively and qualitatively analyze the failure characteristics of the discrete element model under multiple explosive loads. The results show that the method of continuous homology can accurately reflect the topological characteristics of the surrounding rock of the tunnel The maximum one-dimensional bar code connection radius can effectively warn tunnel instability. This provides a new mathematical method for tunnel safety design and disaster prediction research.

**Key words:** Persistent homology; Shallow buried tunnel; Discrete element; Multiple explosive loads

## 1 Introduction

The safety of underground civil air defense engineering, protection engineering and subway tunnel under the action of explosion load is an important issue related to the survival and support ability of underground engineering under the conditions of war or accidental explosion [1]. It is of great theoretical and practical significance to study the sustained coherence characteristics of tunnel structure failure process under explosion load. Because the structure of tunnel surrounding rock is often cut by weak structural planes such as bedding plane and joint surface, its structure presents discontinuity, so scholars mostly use discrete element method to study the damage evolution of tunnel surrounding rock [2-6]. In recent years, persistent homology method has been introduced by scholars into Large Data, Artificial Intelligence, Intelligent Manufacturing, Materials genome project, machine learning and deep learning and other fields [7-13]. At present, there is little research on the persistent homology characteristics of tunnel failure process at home and abroad. However, the topology optimization of tunnel plays an important role in protection and design. Shen Caicai et al. [14] carried out topology optimization analysis on tunnel excavation by introducing topological theory. Lai Hongpeng et al. [15] carried out topology optimization research on lining structure of single arch tunnel and multi arch tunnel with different stress fields based on the structural optimization theory and finite element method of continuum. However, there are some problems in previous studies, such as complex operation and lack of consideration under external force. Persistent homology is a powerful general technology to describe the geometry, structure and topology of data. It can continuously adjust the coverage scale, get the network at all scales, and then

*Corresponding author

E-mail address: 101434@uzz.edu.cn

calculate the change of topological invariants of homology group "from birth to death". Especially, it has the advantage of accurate image processing. Using the images obtained by ground penetrating radar (GPR) in underground engineering, the persistent homology can accurately describe the structural characteristics and evolution law of tunnel rock mass, and can lead the topological information to machine learning and artificial intelligence. In the field of tunnel engineering, it is a new and efficient analysis method. In this article, the dynamic response of surrounding rock is studied by using UDEC software. The topological characteristics of tunnel surrounding rock under multiple explosion loads are analyzed by the method of persistent homology, and the damage evolution law of tunnel surrounding rock is described based on the parameters of dimension bar code.

## 2 Basic principles of persistent homology

Persistent homology is originated from Morsel theory, is a method used to calculate topological features at different spatial resolutions. Persistent homology can detect more continuous features on a wide spatial scale. These features are independent of the filter scale and can better represent the true characteristics of underlying space. It is an efficient and rigorous method to calculate different spatial topological features

### 2.1 Simplicial Complexes

Simplex: Any finite set of vertices, the rank of the maximum independent vector is $n$. If the dimension is $n$, then the simplex $V$ can be expressed as:

$$V = \{v_i, i = 0, ... n\} \quad (1)$$

*A geometric simplicial complex K is a finite set of geometric simplices that satisfy two essential conditions:*
*1. Any face of a simplex from K is also in K.*
*2. The intersection of any two simplices in K is either empty or shares faces.*

### 2.2 Homology group

Suppose $(v_0, v_1, ... v_n)$ is the vertex of $n$-dimensional simplex A, and set the permutation $\phi$, then:

$$(v_0, v_1, ..., v_n) = sign(\phi)\phi(v_0, v_1, ..., v_n) \quad (2)$$

The edge is obtained and generalized to the general case:

$$\partial (v_0, v_1, ..., v_n) = \sum_{i=0}^{n}(-1)^i(v_0, v_1, ..., v_i, ..., v_n) \quad (3)$$

where $(v_0, v_1, ... v_i, ... v_n)$ is the $n$-1-dimensional oriented simplex obtained by removing vertex $v_i$. Let A be a finite simple complex, and let all directed $q$-dimensional simple complexes on a be set $Q = \{q_i, i = 1, ... m\}$, Then, on the set, a free commutative group consisting of linear combinations of integers as coefficients is formed, Recorded as $G_q(A)$.

$$G_q(A) = \left\{\sum_{i=1}^{m} \lambda_i q_0, i \in Z\right\} \quad (4)$$

$G_q(A)$ is called the $q$-dimensional chain group of A, according to the definition of edge, a homomorphism is defined as edge homomorphism, that is:

$$\partial: G_q(A) \rightarrow G_{q-1}(A) \quad (5)$$

Let the edge of a single vertex be zero and $G_{-1}(A) = 0$. Because the kernel of homomorphism $\partial$ is a closed chain group, that is, the elements in the kernel of homomorphism $\partial$ are all closed chains of $G_q(A)$, which is denoted as $Z_q(A)$. Edge closed chain means that a closed chain is the edge of its high one-dimensional chain, Get the chain of $G_q(A)$ by finding the edge of the chain in $G_{q+1}(A)$. The group formed by edge closed chain is called edge closed chain group, which is called $B_q(A)$, then the q-dimensional homology group is defined as:

$$H_q(X,A) = \frac{Z_q(A)}{B_q(A)} \tag{6}$$

**2.3 persistent homology**

For vertex set $V = \{v_0, v_1, ..., v_n\}$, let vertex set be point cloud set, it can be quantified. Now we introduce the concept of connected radius $\varepsilon$ (if the distance between two points is less than $\varepsilon$, then we call the connection between two points as connectivity). Each vertex in $V$ is the center of the circle and the radius is $\varepsilon / 2$. If the intersection of $P$ balls is not empty, then these $p$ vertices constitute a $p$-1-dimensional simple complex. With the increase of $\varepsilon$, the number of simplex forms of the complex increases, this forms a complex flow $A_0 \subset A_1, ..., \subset A_n$, then we find the homology ,groups of complex flow at different times is $H_p(A_0), H_p(A_1), ..., H_p(A_n)$, that is persistent homology.

**2.4 Bar code image**

The bar code plots is used to reflect the topological characteristics of the complex filtration duration of the point cloud set in the process of increasing the connected radius. It is a collection of finite intervals on the real axis R, which can generally be expressed as [a,b] or [a ,+∞], where a,b∈R. The following is a simple example of persistent homology. Figure 1 is the topological invariant Betty number, 0-dimensional Betty number b0 is the number of connected bodies, 1-dimensional Betty number b1 is the number of one-dimensional holes, two-dimensional Betty number b2 is the number of three-dimensional cavities or spaces, and so on, as shown in Fig.1.

|  | point | circle | ball | torus |
|---|---|---|---|---|
| $b_0$: | 1 | 1 | 1 | 1 |
| $b_1$: | 0 | 1 | 0 | 2 |
| $b_2$: | 0 | 0 | 1 | 1 |

**Fig.1 Betty numbers of different dimensions**[16]

Figure 1 shows the Betty number of a point, a circle, an empty sphere and a torus. For a torus, two auxiliary rings are added to explain $b_1 = 2$.

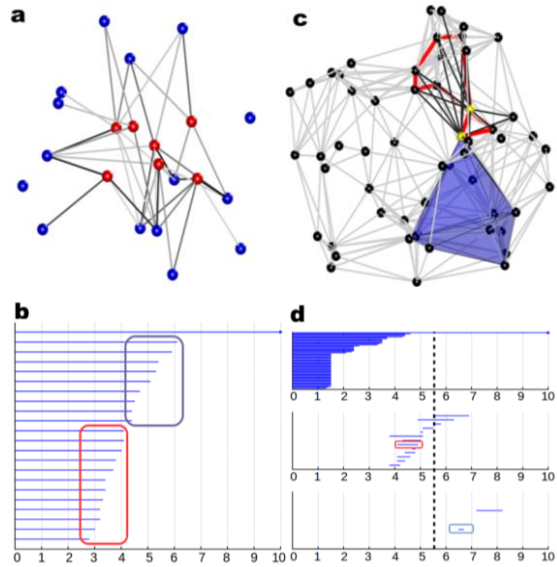

**Fig.2 simple complex and its corresponding barcode**[16]

Figure 2 is a demonstration of the topological characteristics of data from point cloud to bar code graph, (a) represents the points in the space and forms a metric space; (b) is the corresponding b 0 bar code graph. The ordinate is the number of points and the abscissa is the connected radius; (d) is a bar code graph of metric space (c), its upper, middle and lower coordinates are $b_0, b_1, b_2$. When we take different ε, we get the simple complex structure, the first bar code in (d) represents the number of connected bodies, when the $ε$ is small, the ordinate represents the number of its points, but with the increasing of ε, $b_0$ becomes 1, which can be understood as there is only one "point" with the increase of connecting radius; In (d), the second bar code diagram represents the number of one-dimensional holes, and the third bar code diagram represents the number of three-dimensional cavities or spaces, when 0<$ε$<3.82, $b_1$=0,$b_2$=0; when 6.46<$ε$<6.70, one-dimensional hole, a three-dimensional cavity or space is formed, $b_1$=1,$b_2$=1; when $ε$>8.2, $b_1$=0,$b_2$=0.

## 3 Simulation analysis of tunnel stability under multiple explosion loads

### 3.1 Establishment of discrete element model

The discrete element model is established by referring to the actual situation of Chongqing Wukou subway station [17]. The starting and ending point mileage of Wukou tunnel with large-span section breaking normal is dk60 + 920.559-dk61 + 255.889, and the total length of the tunnel is 105.33m. The mining method is adopted for construction. The vault of the tunnel is about 16m from the ground, and the excavation section is 20.40m wide and high After exploration, most of the surrounding rock of the tunnel is sandstone, belonging to grade 4. Considering the influence of tunnel depth and boundary effect on tunnel surrounding rock deformation, and according to the actual engineering data, a discrete element calculation model for this study is established, as shown in Fig.3. The width of model is l=140m, height is $h$=76.5m, top buried depth of tunnel excavation is $f$=16m, section width of tunnel excavation is $l_1$=20.59, height of tunnel excavation is $h_1$=17.94m. The mechanical parameters and joint mechanical parameters of tunnel surrounding rock are shown in Table 1 and table 2. The mechanical parameters and joint mechanical parameters of surrounding rock are selected strictly according to the material parameters corresponding to the surrounding rock grade in the engineering geological report. In the simulation, blasting will cause deformation of rock mass, so Mohr Coulomb criterion is used to set rock mass and joint. According to reference [17], when the joint plane spacing is 5m, the failure area, plastic area and average displacement of surrounding rock are the largest, therefore, this thesis selects two joints with a joint spacing of 5m and an inclination angle of 50°/110° through the surrounding rock of the tunnel. In order to ensure that the model boundary does not affect the accuracy of the simulation, viscous boundary conditions are applied to the bottom and both sides of the model after reaching the equilibrium state under the action of gravity, and the horizontal constraints are applied to both sides, and the vertical constraints are applied to the bottom.

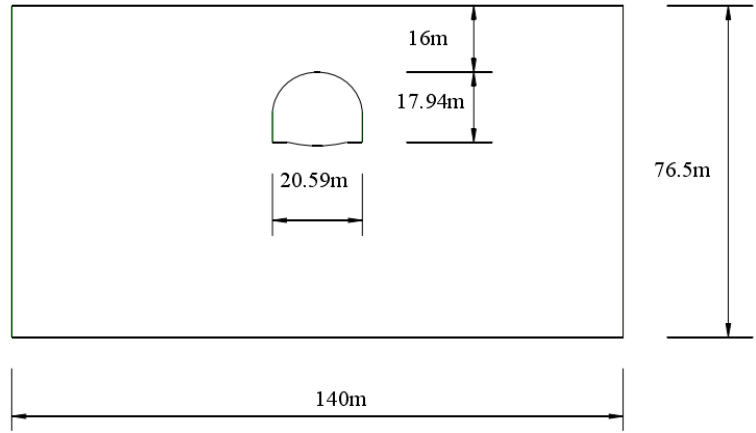

**Fig.3 Model size**

**Tab.1 Mechanical parameters of tunnel surrounding rock**

| Density/kg·m$^3$ | Bulk modulus/GPa | Shear modulus/GPa | Angle of internal friction/° | Cohesive force/MPa | Tensile strength/MPa |
|---|---|---|---|---|---|
| 2300 | 4.17 | 1.92 | 35 | 0.5 | 1.0 |

**Tab.2 Joint mechanics parameters**

| Normal stiffness/GPa | Tangential stiffness/GPa | Angle of internal friction/° | Cohesive force/MPa | Tensile strength/MPa |
|---|---|---|---|---|
| 10 | 10 | 20 | 0.04 | 0 |

The simulation assumes that the ground directly above the tunnel is impacted by explosion, and the specific model uses the explosion load to simulate the effect of the ground explosion on the unsupported tunnel. According to the research in reference [18], when simulating the tunnel under the action of ground explosion load, it is equivalent to applying a uniform load varying with time on the top of the model [18]. In the process of numerical simulation, the time history curve of explosion load is mainly treated as triangle and exponential type [19]. In this thesis, the triangle form is adopted, and the duration of explosion load is generally considered to be $1.0\times10^{-6}$~0.1s [20]. The comprehensive consideration model is shallow tunnel and discrete element calculation. Compared with the calculation method of continuum mechanics, the internal transmission process of stress wave is relatively slow, the duration of explosion load is 35 ms, in which the pressure rise time is 5 ms and the positive pressure action time is 30 ms. At the same time, the value of explosion load is relatively large to simulate the maximum damage.

The explosion load is equivalent uniform load, the calculation formula is as follows:

$$P = \frac{2R}{L} \cdot P_0 \quad (7)$$

Where $P$ is the equivalent uniform load, $R$ is the blast hole radius, and $L$ is the hole spacing. In this thesis, the value of the blast hole spacing in the model is 1.25m, and $P_0$ is the load produced by a single blast hole.

$$P(t) = P_m \cdot f(t) \quad (8)$$

$$P_m = \frac{1}{8} \cdot \rho_c \cdot D^2 \cdot \kappa_c^{-6} \cdot \eta \quad (9)$$

Where $f(t)$ is the exponential function of the explosion load with time, $P_m$ is the peak value of the explosion load, $D$ is the detonation degree of the explosive, $\rho_c$ is the charge density, $K_c$ is the decoupling coefficient, and $\eta$ is the expansion multiple of the hole wall pressure affected by the detonation gas. The value of this model is 9. Therefore, the formula of uniform load is as follows:

$$P = \frac{R}{4L} \cdot \rho_c \cdot D^2 \cdot \kappa_c^{-6} \cdot \eta \cdot f(t) \quad (10)$$

According to the reference [18] No.2 rock parameter, the explosive detonation velocity is 3.5km/s, the charge density is 1000 kg/m$^3$, the blast hole radius is 0.055m, and the charge uncoupling coefficient is 1.447. After calculation, $P_m=1.5 \times 10^9$ Pa, $P=1.38 \times 10^8 f(t)$.

## 3.2 Dynamic response of tunnel under multiple explosion loads

Under the natural state of gravity and initial boundary conditions, the mechanical state of tunnel surrounding rock tends to be stable. The maximum displacement of the model and the displacement of the collapse area of the upper part of the tunnel are 0.232m and 0.208m respectively. After applying the first to the fifth explosion load on the model, the vector and cloud images of velocity and displacement are analyzed, and the following four tables are obtained (Tab.3, Tab.4, Tab.5,Tab,6)：

**Tab.3 Maximum displacement of model**

| burst/times | 1 | 2 | 3 | 4 | 5 |
|---|---|---|---|---|---|
| displacement(m) | 0.389 | 1.149 | 1.997 | 3.028 | 4.758 |

**Tab.4 Model maximum speed**

| burst/times | 1 | 2 | 3 | 4 | 5 |
|---|---|---|---|---|---|
| speed(m/s) | 17.42 | 20.76 | 27.07 | 38.87 | 46.47 |

**Tab.5 Displacement of the upper collapse zone of the tunnel**

| burst/times | 1 | 2 | 3 | 4 | 5 |
|---|---|---|---|---|---|
| displacement(m) | 0.272 | 0.804 | 1.389 | 2.121 | 3.331 |

**Tab.6 Speed of collapse zone in upper tunnel**

| burst/times | 1 | 2 | 3 | 4 | 5 |
|---|---|---|---|---|---|
| speed(m/s) | 10.45 | 14.54 | 16.24 | 23.32 | 27.88 |

It can be seen from the table of the maximum displacement of the model (Tab.3) and the displacement table of the collapse area at the upper part of the tunnel (Tab.5). It can be seen that with the increase of blasting times, the increase of displacement is more and more large. After the first blasting load was applied, the descending displacement of the top part of the tunnel increased by 0.064m and the maximum displacement of the tunnel increased by 0.157m compared with the displacement under natural state. After the first explosion, the descending speed of the collapse area at the upper part of the tunnel accelerates. With the increase of the number of explosions, the descending displacement of the top part of the tunnel increases by 0.596 m, 1.181 m, 1.913 m and 3.123 m, respectively. However, compared with table 3 and table 5, it can be found that the upper part of the tunnel is not the place with the maximum collapse displacement, and the maximum displacement of the tunnel is on the left and right sides of the bottom of the tunnel .The left and right bottom of the surrounding rock of the road is also the largest stress position. By analyzing the maximum velocity table of the model (Tab.4) and the velocity table of the collapse area at the upper part of the tunnel (Tab.6), we can know that the velocity is relatively large because the explosion load is relatively large, but the actual model is a shallow tunnel. In the first explosion, the maximum speed of the tunnel is 17.42m/s, and the maximum speed of the collapse area at the upper part of the tunnel is 10.45m/s. With the increase of the number of explosions, the speed of the surrounding rock of the tunnel is also faster and faster. After the first blasting, the speed of the rock mass has a very large increase, which shows that the more the number of blasting, the more obvious the damage effect on the stability of rock mass 。 However, in view of the fact that the displacement characteristics of surrounding rock block are mainly selected in this thesis, and the failure speed of explosion in tunnel surrounding rock is also related to the size of explosion load and mechanical parameters of surrounding rock, so it is not studied in depth.

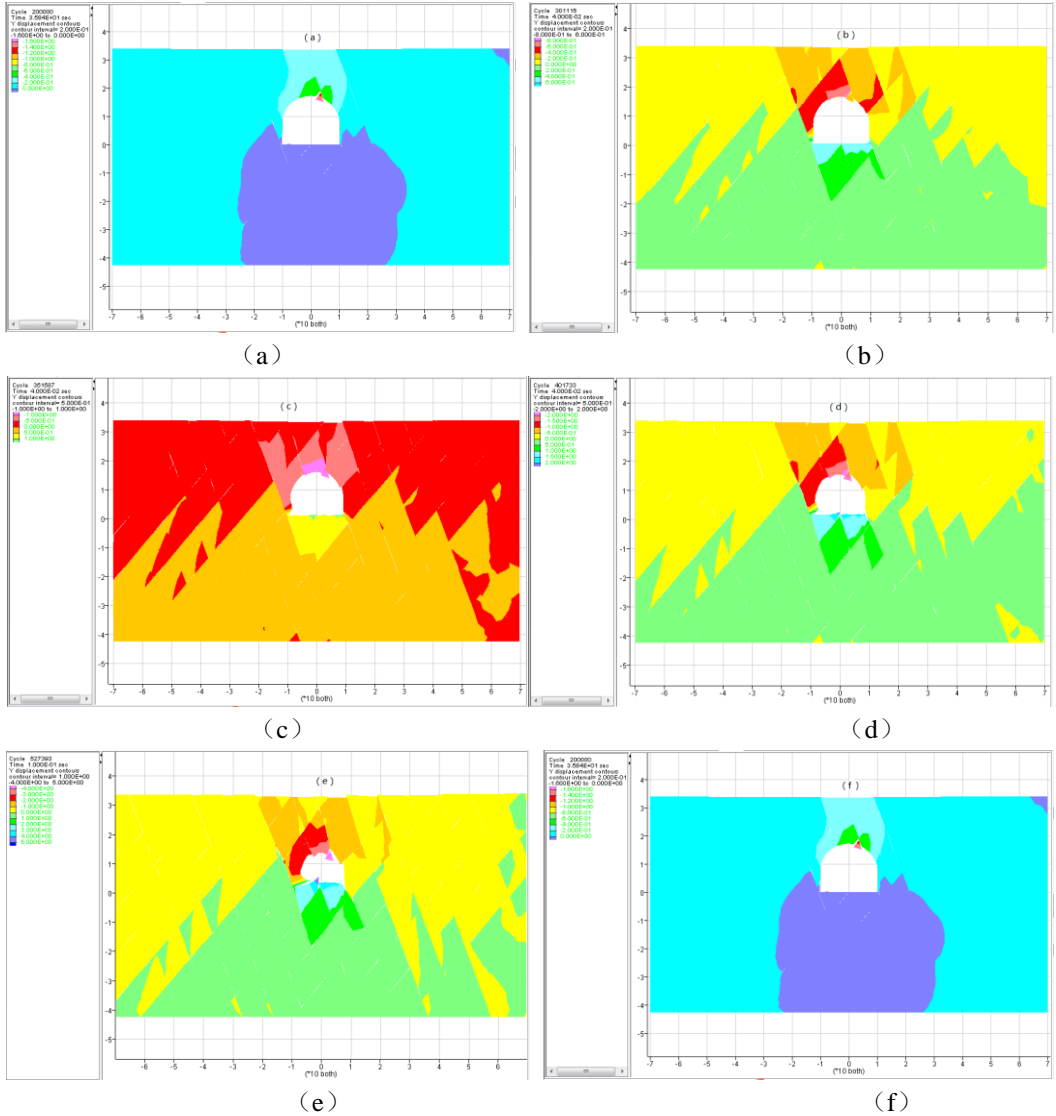

**Fig.4 Displacement cloud images of one to five explosions and tunnel under natural conditions**

Fig.4 (a ~ f) shows the tunnel displacement nephogram after 1 to 5 times of explosion load and under natural state. When the foundation soil and tunnel structure are fixed, the maximum displacement of tunnel structure increases with the increase of explosion load intensity and time, and decreases with the increase of tunnel buried depth, showing an approximate linear relationship [21]. Combined with the factors in this thesis, such as large and multiple blasting loads, no support and reinforcement of tunnel surrounding rock, and sandstone characteristics of rock mass, it can be seen that the tunnel will produce large deformation. It can be seen from the figure that the tunnel has collapsed seriously after five times of explosive loading. According to the research of Zheng Yingren et al. [22], it is found that the strength reduction method has been applied to the judgment of tunnel stability. After calculating the overall safety factor of the tunnel surrounding rock, the potential slip surface is obtained, which provides a reliable basis for the subsequent tunnel stability evaluation. The tunnel model studied in this thesis is a shallow tunnel, so the main failure mode is slip.

For the Mohr Coulomb model which studied in this article, when the structural plane does not slide or open, the tensile strength and shear strength are:

$$\left. \begin{array}{l} T_{max} = -TA_c \\ F^t_{max} = cA_c + F^n \tan\phi \end{array} \right\} \quad (11)$$

Where $T_{max}$ is the maximum tensile force, $T$ is the tensile strength of the structural plane, $A_c$ is the contact area, $F^t_{max}$ is the shear strength of the structural plane, $F^n$ is the normal stress, $c$ and $\phi$ are the cohesion and internal friction angle of the structural plane respectively.

When the surrounding rock of the tunnel is sliding, its tensile strength and shear strength are:

$$\left. \begin{array}{l} T_{max} = T_{residual} \\ F^n_{max} = C_{residual}A_c + F^n \tan\phi \end{array} \right\} \quad (12)$$

Where, $T_{reaidual}$ is the residual tension, $F^n_{max}$ is the residual shear, $C_{residual}$ is the residual cohesion, $A_c$ is the contact area.

UDEC program can reduce the strength parameters of deformation body and structural plane. When the tunnel is unstable, the built-in strength reduction calculation program can calculate the overall safety factor of the tunnel. According to the calculation, when the reduction coefficient is 1.52, the tunnel is in the limit state.

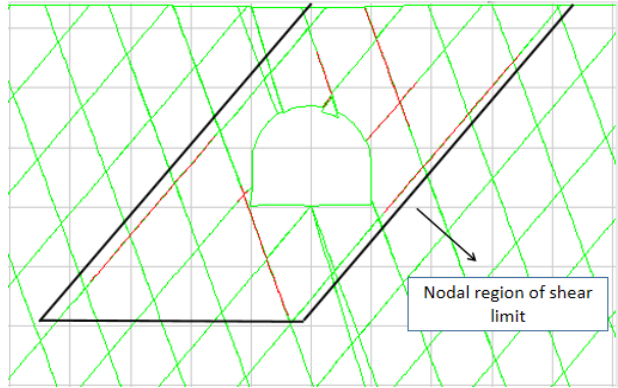

**Fig.5 Node graph of shear limit after the first blasting**

It can be seen from the node graph (Fig. 5) that the boundary of the collapse area has reached the shear strength limit after the first blasting. Fig. 4 displacement nephogram after each explosion shows that after the first explosion load is applied, the collapse of surrounding rock of tunnel becomes more and more obvious with the increase of explosion times. After the first explosion load, the collapse area has reached the shear strength limit, and then a large area of collapse occurs.

## 4 Persistent homology characteristics of tunnel surrounding rock under multiple explosion loads

The mathematical method of persistent homology is introduced to analyze the stability of tunnel surrounding rock under multiple explosion loads. When studying the block displacement of tunnel surrounding rock under the action of explosion load, we first divide the surrounding rock into images composed of numerous blocks, and we regard the block as a point cloud in European space. Figure 6 is the scatter diagram of the surrounding rock block of the tunnel in the natural state of excavation, that is, the point cloud of the surrounding rock of the tunnel under the natural state.

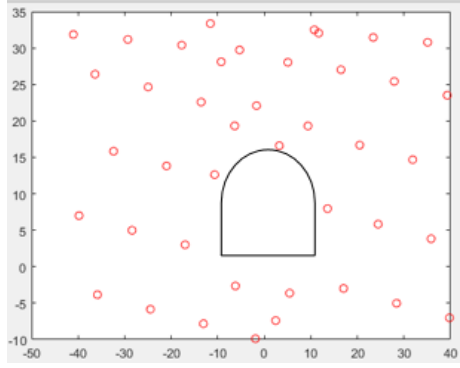

**Fig.6 Scattered Point Map of Tunnel Block in Natural State**

42 blocks which around the surrounding rock of the tunnel and whose displacement is greater than 0.01m under the explosion load are selected as the point clouds, and the relevant parameters of the persistent homology topological characteristics are obtained by using javeplex software.

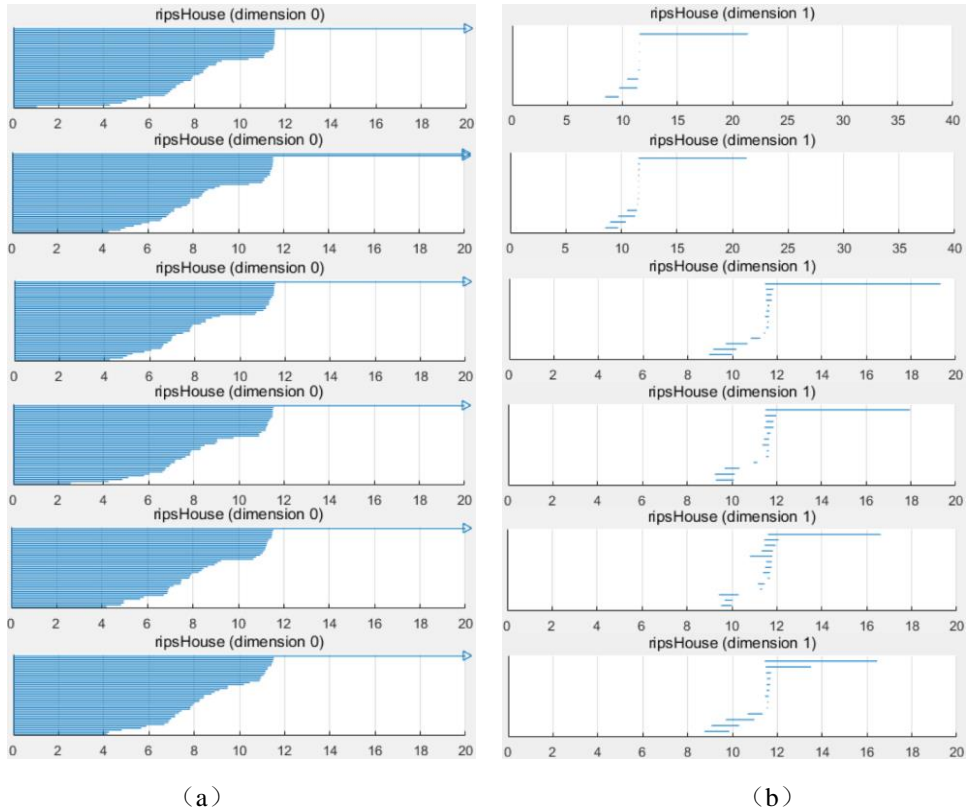

(a)                                     (b)

**Fig.7 Natural state and 1 to 5 times of exploding tunnel surrounding rock block $b_0 / b_1$ barcode**

Fig. 7 (a) is a bar code graph of 0-dimensional Betty number. It can be seen from the graph that most of the connecting radius between blocks is in the range of 0 ~ 12. From the natural state to 1 to 5 explosions, the maximum value of ε has little change. When the selected block area is fixed, the displacement of the edge block is relatively small due to the size of the explosion load. When ε is less than 1, B0 from excavation to the fifth explosion is 42. At this time, from natural state to 1-5 explosion, due to the relatively small connection radius, the number of vertical coordinates also shows the number of blocks selected in the tunnel space; When 1 < ε < 11.65, with the increase of blasting times, the upper layer of 0-dimensional Betty number shows a trend of "thinning" and decreasing, which indicates that the spacing of tunnel surrounding rock blocks is narrowing, which is due to the increase of blasting times, which causes the tunnel surrounding rock to collapse to the middle; When 11.65 < ε, all the blocks are connected to form a connected body. However, from the beginning of the explosion, the ε began to decrease gradually, which is because the surrounding rock of the tunnel began to collapse and the "hole" of the tunnel began to be compressed. At this time, part of the ε began to decrease. It can be seen from the 1-D Betty number bar code in Fig. 7 (b), with the increase of explosion times, ε always changes in an interval, $8 < \varepsilon < 25$, with the natural state to multiple times of explosive loading, the longest straight line ε of 1D barcode gradually decreases, which indicates that the diameter of 1D "holes" formed between blocks is gradually decreasing. This "hole" is caused by the position relationship of blocks or their geometric structure in European space. Of course, there are not only one "hole", but also many small "holes", which belongs to the bar code noise, we can ignore it.

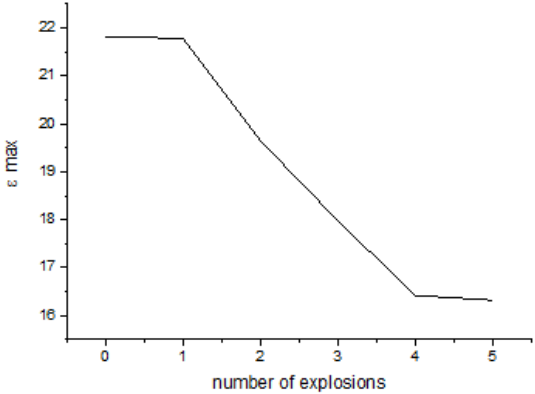

**Fig.8 One-dimensional bar code graph $\varepsilon$ maximum change with the number of explosions**

Fig. 8 is the curve of the maximum value of one-dimensional bar code ε in Fig. 7 (b) from the natural state to the application of explosive load for 1 to 5 times. It can be seen from the figure that the maximum value of ε decreases with the increase of explosion times, and the maximum value of ε under natural state is 21.82. With the increase of blasting times, the collapse of tunnel is becoming more and more serious, and the tunnel is shrinking. After the first explosion, the maximum value of ε is 21.78, and the variation is 0.04. The maximum value after 2-5 times of blasting is 2.17, 3.84, 5.4 and 5.7 respectively, which indicates that with the increase of blasting times, the connecting radius of holes is decreasing. After the first explosion load is applied, the change of ε is small. When the second explosion load is applied, compared with the first explosion load, the change of ε is 53 times. According to Fig. 4, Fig. 5 and Fig. 8, after the first blasting load is applied, the surrounding rock of the tunnel reaches the shear limit, that is, the tunnel begins to collapse violently. After the collapse area reaches the shear limit, the holes formed between blocks begin to shrink, and the maximum value of ε changes rapidly and becomes smaller and smaller. At this time, it can be seen that the shear limit of tunnel surrounding rock can be established with the maximum value of ε in one-dimensional bar code diagram. When the surrounding rock of the tunnel breaks through the shear limit under the explosion load, the tunnel begins to collapse, and the maximum value of ε begins to decrease, and the degree of reduction is linear with the collapse speed. With the increase of explosion times, the displacement of collapse increases, and the maximum "hole" formed between blocks is reduced. According to the above inference, the collapse area of tunnel surrounding rock reaches the limit of shear strength after the first explosion load, and then ε = 21.78. When ε is less than 21.78, the tunnel will be damaged. If the explosion load is applied to the tunnel, the maximum value of ε will decrease rapidly and the damage of surrounding rock will increase rapidly. It can be seen that the safety factor of tunnel surrounding rock can be expressed by the maximum value of ε in one-dimensional bar code diagram. When the maximum value of ε exceeds the limit value, the tunnel begins to lose stability and the maximum value of ε begins to decrease.

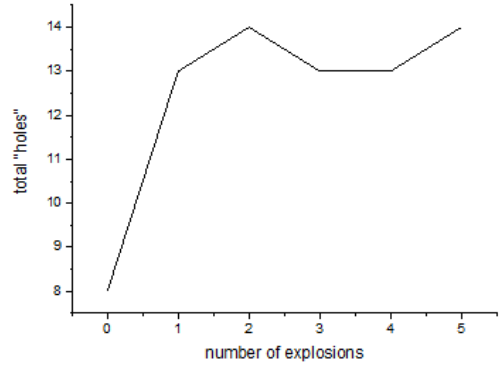

**Fig.9 Total number of "holes" formed during explosion**

Fig.9 shows the number of ordinates of one-dimensional bar code diagram under natural state and after one to five times of blasting, i.e. the total horizontal line of Fig.7 (b). Fig.9 shows the number of "holes" continuously formed in the surrounding rock of the tunnel during the process of increasing the connecting radius. In natural state, the total number of "holes" in 1D barcode map is 8, and 8 "holes" are formed. At this time, the redundant holes are caused by removing some holes with small displacement and small area in calculation. After the explosion load was applied, it increased rapidly to 13 and then changed from 13 to 14. Under the explosion load, the tunnel will be destroyed, and the surface protruding block and internal block displacement will be formed, and the number of "holes" will increase rapidly with the application of explosion load, that is to say, the number of "holes" will increase when the explosion load is applied, and the overall trend is increasing.

## 5 Conclusion

A new mathematical method, persistent homology, is introduced into the research field of tunnel safety design and disaster prediction. The feasibility and accuracy of applying persistent homology to tunnel engineering are proved by studying the failure characteristics of a tunnel model under multiple explosion loads

Based on the discrete element method (DEM), the dynamic response and the damage evolution process of tunnel surrounding rock are obtained after each explosion load. The results show that: with the increase of blasting times, the 0-dimensional Betty number changes a little, that is, blasting changes the distribution of surrounding rock structure, and the overall 0-dimensional $\varepsilon$ decreases, which reflects that the blasting makes the surrounding rock collapse close, and from the "hole" number of 1-dimensional Betty number, it can be seen that the explosion load will increase the number of "holes", and the overall trend is increasing; the safety factor of tunnel surrounding rock can be used. When the maximum value of $\varepsilon$ in one-dimensional bar code diagram breaks through the limit value, the tunnel begins to lose stability and the maximum value of $\varepsilon$ also begins to decrease. The maximum value of $\varepsilon$ in one-dimensional bar code diagram can be used to judge whether the tunnel surrounding rock reaches the maximum shear strength limit, so as to determine the zero point of tunnel surrounding rock collapse

Using the method of persistent homology to study shallow buried tunnel under explosion load can be combined with the current research in tunnel engineering field and ground penetrating radar and other means to obtain the geometric structure of underground tunnel engineering, that is, the topological characteristics of the tunnel can be analyzed by the method of persistent homology, and the data can be used in machine learning fields such as support vector machine and BP neural network. It provides a new idea for the study of tunnel design and safety.

## Acknowledgement

This work was supported by the Project Funded by Jiangxi Provincial Department of Science and Technology (No. 20192BBEL50028).